\documentclass[conference]{IEEEtran}

\usepackage{cite}
\usepackage{listings}
\usepackage{multirow}
\usepackage{booktabs}
\usepackage{graphicx}
\graphicspath{{./images/}}
\usepackage{hyperref}
\hypersetup{colorlinks=true,linkcolor=black,urlcolor=black,citecolor=black}
\usepackage{xcolor}
\usepackage{subcaption}
\usepackage{comment}

\begin{document}

\begin{titlepage}
\quad\\[1cm]
\makeatother
	{\Huge IEEE Copyright Notice}\\[0.5cm]
	{\large \copyright \ 2025 IEEE. Personal use of this material is permitted. Permission from IEEE must be obtained for all other uses, in any current or future media, including reprinting/republishing this material for advertising or promotional purposes, creating new collective works, for resale or redistribution to servers or lists, or reuse of any copyrighted component of this work in other works. \\[0.5cm]}    
    {\large Cite as:\\[0.1cm]}
    {\large L. Perugini, A. Vesco, “An Efficient TLS 1.3 Handshake Protocol with VC Certificate Type,” in \textit{Proceedings of IEEE 22\textsuperscript{nd} Consumer Communications Networking Conference (CCNC)}, Las Vegas, Nevada, USA, 2025, pp. 1–9, doi:10.1109/CCNC54725.2025.10975914.}
\end{titlepage}

\title{An Efficient TLS 1.3 Handshake Protocol with\\ VC Certificate Type}

\author{\IEEEauthorblockN{Leonardo Perugini and Andrea Vesco}
\IEEEauthorblockA{Cybersecurity Research Group -- LINKS Foundation\\
Via P.C. Boggio, 61 -- Torino 10138, Italy\\
Email: \{leonardo.perugini, andrea.vesco\}@linksfoundation.com}}

\maketitle

\begin{abstract}
The paper presents a step forward in the design and implementation of a Transport Layer Security (TLS) handshake protocol that enables the use of Verifiable Credential (VC) while maintaining full compliance with RFC-8446 and preserving all the security features of TLS 1.3. The improvement over our previous work lies in the handshake design, which now only uses messages already defined for TLS 1.3. The design has an incredibly positive impact on the implementation, as we made minimal changes to the OpenSSL library and relied mostly on a novel external provider to handle VC and Decentralized IDentifier (DID) related operations. The experimental results prove the feasibility of the design and show comparable performance to the original solution based on Public Key Infrastructure (PKI) and X.509 certificates. These results pave the way for the adoption of Self-Sovereign Identity (SSI) in large-scale Internet of Things (IoT) systems, with a clear benefit in terms of reducing the cost of identity management. 
\end{abstract}

\IEEEpeerreviewmaketitle

\section{Introduction}
\label{sec:intro}

The number of Internet of Things (IoT) devices is multiplying every year. According to Statista \cite{statista-iot}, the number of IoT nodes is expected to grow to 29 billion by 2030. At the same time, IoT systems, with their poor security reputation, have become an entry point for adversaries to target critical IoT systems. All leading cybersecurity analysts report a steady increase in attacks against IoT nodes from year to year. Providing these devices with secure digital identities is a critical first step in keeping pace with evolving threats. Especially in the case of large-scale IoT systems, automation is crucial for provisioning, updating and revoking node identities. Current solutions rely on a centralized Public Key Infrastructure (PKI) with X.509 certificates \cite{rfc5280}. However, this process is largely seen by industry partners as inefficient and costly. The use of PKI and X.509 certificates requires human intervention, as there is no standard automated solution tailored to IoT nodes and protocols. For instance, anytime the identity of a node changes or its public key is rotated, the X.509 certificate must be revoked, and a new one must be issued. The revoked certificate must be added to a certificate revocation list, which is not immediately propagated throughout the IoT system. On top of this, the Certificate Authorities (CAs) come with other drawbacks. They impose a cost on the system administrator to add a trusted third party into trust relationships, and they are vulnerable to a single point of failure, as a single breach can compromise all issued certificates, requiring an update of all node identities before network operations can resume. ~\cite{8717910,10230166,9705036,10076881,2020101658}.

The Self-Sovereign Identity (SSI) model~\cite{SSIbook} represents an alternative solution to overcome the limitations of the centralized PKI and reduce the complexity and the related cost of X.509 certificate management in large scale IoT systems. SSI is an emerging decentralized digital identity model. It gives a node full control over the data it uses to generate and prove its identity.
An SSI native IoT node generates its identity key pair and stores the public key in a distributed ledger for other nodes to authenticate it. A node's Decentralized IDentifier (DID) represents the address in the distributed ledger where other nodes can retrieve its public key. Once these two components have been generated, a node can request a Verifiable Credential (VC) from the Issuer in the IoT system. The VC contains the metadata and claims about the identity of the node that holds it. 
The combination of the key pair, the DID and at least one VC shapes the digital identity in an SSI native IoT system. The composition of the digital identity reflects the decentralized nature of the SSI model. In fact, the distributed ledger overcomes the single point of failure limitation of the traditional PKI, as no single party is in control of the ledger that contains all public keys. Plus, the separation of public key information (stored in the ledger) from identity attributes (stored in VCs) also comes with several advantages: (\textit{i}) a node can rotate its key pair without the need of refreshing the VC, by updating autonomously the information in the ledger, (\textit{ii}) a node can immediately revoke its DID if the key pair gets compromised invalidating right away the VC, (\textit{iii}) the identity life cycle of a device does not require human intervention, and (\textit{iv}) the scale at which identity can be provided to devices increases dramatically.

Given these benefits, our goal is to contribute to the adoption of SSI in IoT systems and its use for authentication purposes. However, most of the discussions and proposals for the adoption of the SSI model consider an implementation at the application layer of the TCP/IP stack \cite{ssi-iot, diam-iot}. The problem with application layer authentication is, first, the lack of established protocols that can be used in IoT systems and, second, the need to use two different technologies simultaneously, X.509 certificates on the server side to establish a secure channel and VCs on the client side to complete the mutual authentication on top of the established secure channel. This creates an overall friction in the adoption of the SSI model. 

We believe in implementing (mutual) authentication at the transport layer of the TCP/IP stack directly within the Transport Layer Security (TLS) protocol \cite{rfc8446} with VC. Using VC in TLS maximizes the benefits of adopting SSI while avoiding the use of PKI and X.509 certificates and their associated costs. In addition, performing authentication at a lower layer in the stack reduces the attack surface and avoids potential vulnerabilities that the application layer may add. 

Enabling the use of VCs requires extending the TLS 1.3 handshake protocol to work with VCs in addition to the existing certificate types (\textit{e.g.}, X.509, raw public key) to ensure interoperability. We have already proposed an extension of the TLS 1.3 handshake protocol in \cite{PERUGINI2024101103} enabling the use of VCs and DIDs while preserving all security features of the original handshake. The design has been validated with a real implementation in OpenSSL and statistically relevant experimental results in real-world settings. The design in \cite{PERUGINI2024101103} enables the use of VC and DID for authentication by adding an extension and two new messages, fully compliant with RFC-8446 \cite{rfc8446}. Despite full compliance and interoperability our design required a modification to the state machine of the protocol implementation. There is a high cost associated to the modification and associated security review of the TLS handshake state machine. This paper presents a new efficient design of the TLS 1.3 handshake protocol that minimizes the changes with respect to the original one hence make it more appealing for a possible standardization \cite{vesco-vcauthtls-01} and upstream contribution to the OpenSSL library.

While working on the design and implementation of the new efficient handshake protocol, an interesting work appeared in literature \cite{did-link}. The authors acknowledge our previous work \cite{PERUGINI2024101103, claudio2023} and they have the same goal of performing authentication in the TLS with VCs and DIDs. Their design divides the overall authentication process in two phases: the pseudo-anonymous authentication with DIDs in the TLS handshake protocol and the identification with Verifiable Presentations (VPs) in the TLS record protocol. However, to leave the handshake unchanged they wrap DIDs inside self-signed X.509 certificates, but this implies the overhead of the certificate generation and management, and the exchange of unnecessary data. In addition, their overall (mutual) authentication process requires a higher number of round trip times (RTTs) than our solutions, because of the identification phase. Furthermore, the pseudo-anonymous authentication with DIDs establish the secure channel even though the identity of the endpoints is still unverified. In a mutual authentication scenario, any adversary can create its own DID and establish the secure channel with the only purpose of draining resources on server side. In our solutions an endpoint never finalizes the handshakes with an unidentified endpoint. 
 
This paper present our new design of an efficient TLS handshake protocol. It does not add any new message or delete any existing one from the original handshake, but only embeds a new extension and adds the \texttt{VC} certificate type to the list of accepted values in the \texttt{client\_certificate\_type} and \texttt{server\_certificate\_type} defined in \cite{rfc7250}. These two extensions provide by default hybrid handshakes and a fallback mechanism to X.509 certificates in case other certificate types are not supported. This novel approach leaves untouched the core of the handshake protocol, and it is meant to ease adoption. 
We implemented most of the additional functionalities in an external provider for OpenSSL.

\section{Self-Sovereign Identity}
\label{sec:ssi}

The SSI reference model consists of three layers. Each layer contributes to the generation of the identity and defines the basic principles for trustable interactions with the other nodes. Note that the SSI model subtends the peer-to-peer relationship between nodes in the system. 

Layer 1 is implemented by any Distributed Ledger Technology (DLT) acting as a Root of Trust (RoT) for public identity data. DLTs are distributed and immutable means of storage by design~\cite{DLTs}. The Decentralized IDentifier (DID)~\cite{DID} is the new type of globally unique identifier designed to verify a node. The DID is a Uniform Resource Identifier (URI) in the form
\begin{verbatim}
did:method-name:method-specific-id
\end{verbatim}
where \verb+method-name+ is the name of the DID method used to interact with the DLT and \verb+method-specific-id+ is the pointer to the DID Document stored in the distributed ledger. Thus, DIDs associate a node with a DID Document~\cite{DID} to enable trusted interactions with it. Here is an example of a DID Document \cite{DID} containing the DID and a verification method with an Ed25519 public key used by the node for authentication purposes

\begin{footnotesize}
\begin{verbatim}
"id": "did:method-name:method-specific-id",
"authentication": [{
  "id": "did:method-name:method-specific-id#keys-1",
  "type": ".. type of verifiation method ..",
  "controller": "did:method-name:method-specific-id",
  "publicKeyJwk": ".. JWK encoded public key .."
}] 
\end{verbatim}
\end{footnotesize}

The DID method~\cite{DID,DID-registry} is a software implementation used by a node to interact with the DLT of choice. In accordance with W3C recommendation~\cite{DID}, a DID method provides the functionalities to
\begin{itemize}
    \item \textbf{Create} a DID: generate an identity key pair ($sk, pk$) for authentication purposes, the corresponding DID Document containing the public key $pk$ and store the DID Document in the distributed ledger at the \verb+method-specific-id+ pointed to by the DID, 
    \item \textbf{Resolve} a DID: retrieve the DID Document from the \verb+method-specific-id+ on the ledger pointed to by the DID, 
    \item \textbf{Update} a DID: generate a new key pair ($sk', pk'$) and store a new DID Document to the same or a new \verb+method-specific-id+ if the node needs to change the DID, and 
    \item \textbf{Deactivate} a DID: provide an immutable evidence in the ledger that the DID has been revoked by the owner. 
\end{itemize}
The DID method implementation is ledger-specific and makes the upper layers independent of the DLT of choice.

Layer 2 uses DIDs and DID Documents to establish a cryptographic trust between two nodes. In principle, both nodes prove the ownership of their private key $sk$ bound to the public key $pk$ in their DID Document stored in the ledger.

While Layer 2 uses DIDs to start authentication, Layer 3 completes it and also deals with authorization to services/resources using Verifiable Credentials (VCs)~\cite{VC}. A VC is a digital credential that contains additional characteristics of a node's identity beyond its identity key pair, the DID, and the DID Document. 

Layer 3 works in accordance with the classical Triangle-of-Trust. Three different roles coexist:
\begin{itemize}
  \item \textbf{Holder} is the node that owns one or more VCs and generates a Verifiable Presentation (VP) to request access to services/resources from a Verifier. A Holder builds the VP as an envelope of the VC and signs it with its identity private key $sk$;
  \item \textbf{Issuer(s)} is the node that asserts claims about the identity of a node, creates a VC from those claims, and signs it before issuing the VC to the Holder. Issuer(s) is also responsible for revoking VCs for cryptographic integrity and/or for status change purposes~\cite{VC};
  \item \textbf{Verifier} is the node that receives a VP from the Holder and verifies two signatures, one made by the Issuer on the VC and one computed by the Holder on the VP, before granting or denying the access to a service/resource based on the claims in the VC. Verifier has an implicit trust in the Issuer(s).
\end{itemize}

The VC contains the metadata to describe properties of the credential (\emph{e.g.}, \verb+context+, \verb+id+, \verb+type+, \verb+issuer+, \verb+issuanceDate+ and \verb+expirationDate+) and most importantly, the DID and the claims about the identity of the node in the \verb+credentialSubject+ field.

The following is an example of VC~\cite{VC} of type \verb+IoTCredential+ for an IoT node issued and signed by the Issuer identified by its DID \verb+did:method-name:abcdefghi+

\begin{footnotesize}
\begin{verbatim}
"@context": 
    ["https://www.w3.org/2018/credentials/v1"],
"id": "https://address/credentials/1",
"type": ["VerifiableCredential", "IoTCredential"],
"issuer": "did:method-name:abcdefghi",
"issuanceDate": ".. date and time ..",
"expirationDate": ".. date and time ..",
"credentialSubject": {
    "id": "did:method-name:123456789",
    .. properties to describe the identity ..
},
"proof": {
    "type": "DataIntegrityProof",
    "cryptosuite": "eddsa-rdfc-2022",
    "created": ".. date and time ..",
    "proofPurpose":	"assertionMethod",
    "verificationMethod": "did:method-name:abcdef
      ghi#key-1",
    "proofValue": ".. Issuer's signature .."
} 
\end{verbatim}
\end{footnotesize}

\section{Transport Layer Security}

The TLS protocol\cite{rfc8446} is meant to establish a secure communication channel between a client and a server over the Internet. The secure channel provides server (and optionally client) authentication, confidentiality and integrity of messages in transit. 

\subsection{Original handshake}
\label{sec:original-handshake}

\begin{figure*}[t]
    \begin{subfigure}{0.45\textwidth}   
        \centerline{\includegraphics[width=\columnwidth]{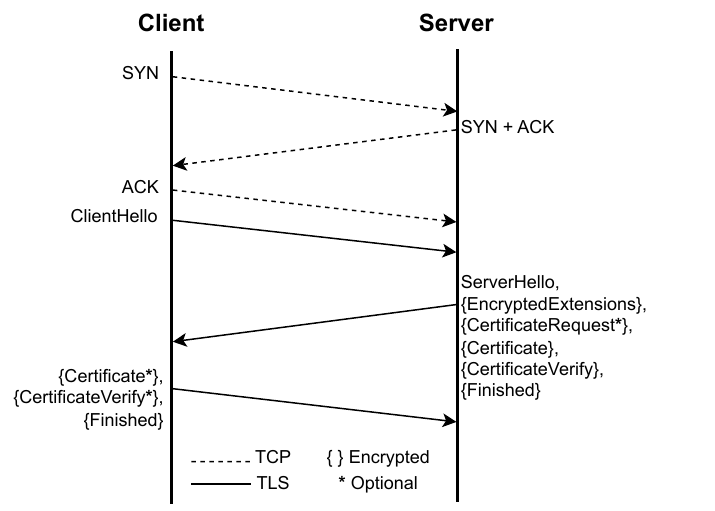}}
        \caption{}
        \label{fig:original-tls-hs}
    \end{subfigure}
    \begin{subfigure}{0.5\textwidth}
        \centerline{\includegraphics[width=\columnwidth]{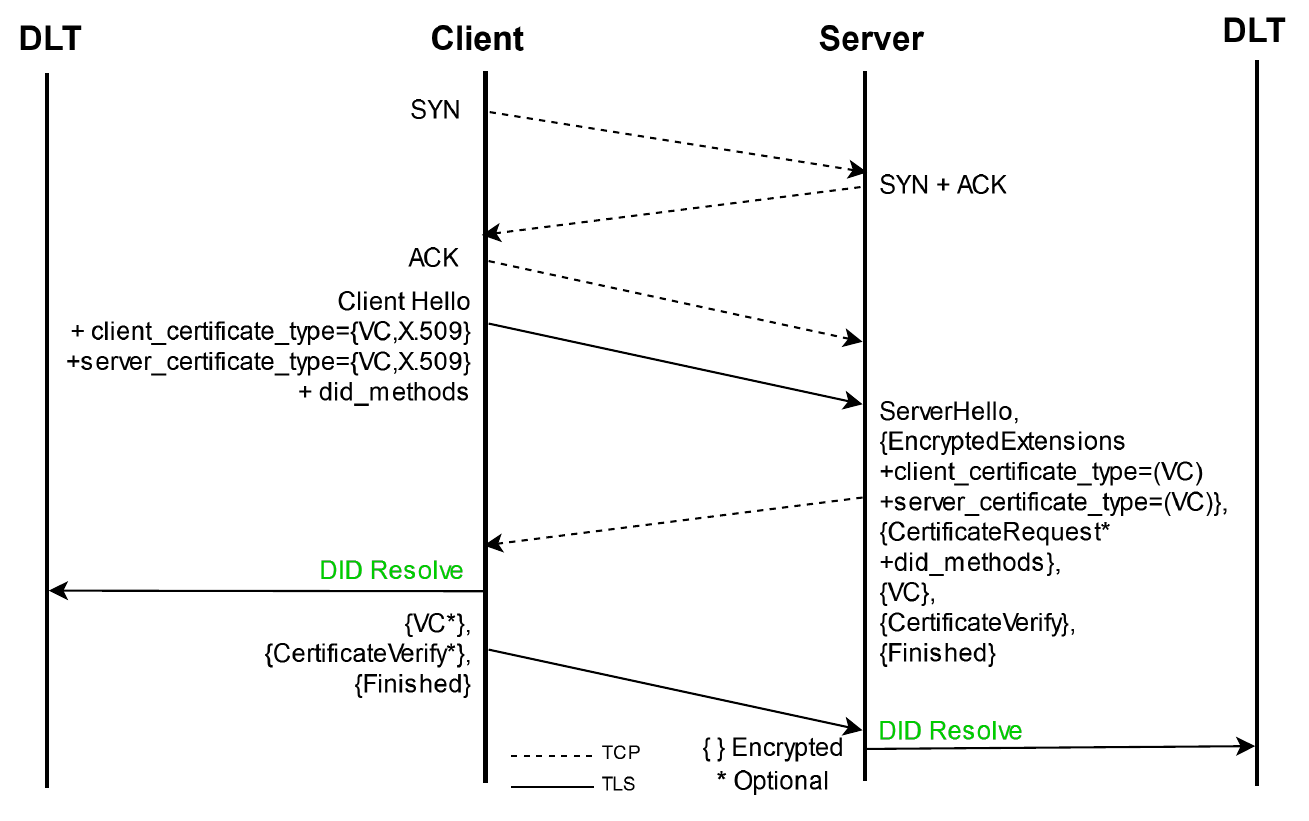}}
        \caption{}
        \label{fig:vc-tls-hs}
    \end{subfigure}
\caption{Flow of messages in (a) the original TLS 1.3 handshake protocol and (b) the TLS 1.3 handshake protocol with \texttt{VC} certificate type.}
\label{fig:original-and-ssi-hs}
\end{figure*}

The TLS 1.3 handshake protocol establishes the secure channel by exchanging the messages shown in Fig.~\ref{fig:original-tls-hs}. 
Upon the handshake, client and server negotiate cryptographic parameters through the exchange of \texttt{ClientHello} and \texttt{ServerHello} messages. Those parameters are the symmetric cipher and hash algorithm used to ensure confidentiality and integrity, respectively. 
The client and the server generate a secret by using ephemeral Diffie–Hellman (EDH) key exchange and use the HMAC-based Extract-and-Expand Key Derivation Function (HKDF) algorithm with the negotiated hash algorithm to derive the session keys from the secret. 
Then, the server authenticates with the client by sending the \texttt{Certificate} message containing the certificate chain (Root CA certificate excluded), and the \texttt{CertificateVerify} message which is the signature over all the exchanged messages computed with its private key. Finally, the server sends the \texttt{Finished} message to ensure the integrity of the handshake. It calculates a {Hash-Based Message Authentication Code (HMAC)} of all the messages exchanged, employing the negotiated hash algorithm, and a {Message Authentication Code (MAC)} key derived from the session key.
The client verifies the validity of the certificate chain, the signature in the \texttt{CertificateVerify} message to check the identity of the server and the HMAC contained in the \texttt{Finished} message.
The server can request client authentication by sending the \texttt{CertificateRequest} message. The client authenticates with the server in the same way.

\begin{figure*}[t]
    \begin{subfigure}{0.55\textwidth}   
        \centerline{\includegraphics[width=\columnwidth]{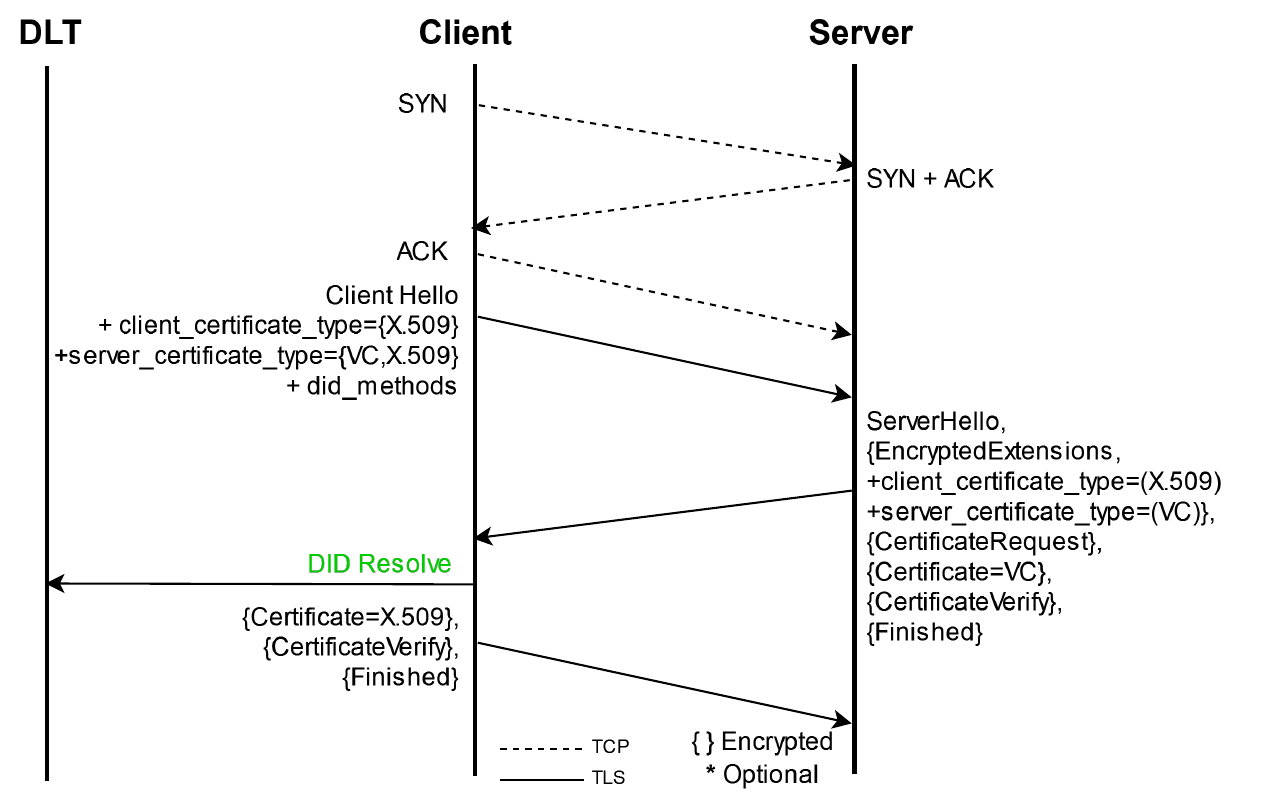}}
        \caption{}
        \label{fig:client-x509-server-vc-hs}
    \end{subfigure}
    \begin{subfigure}{0.5\textwidth}
        \centerline{\includegraphics[width=\columnwidth]{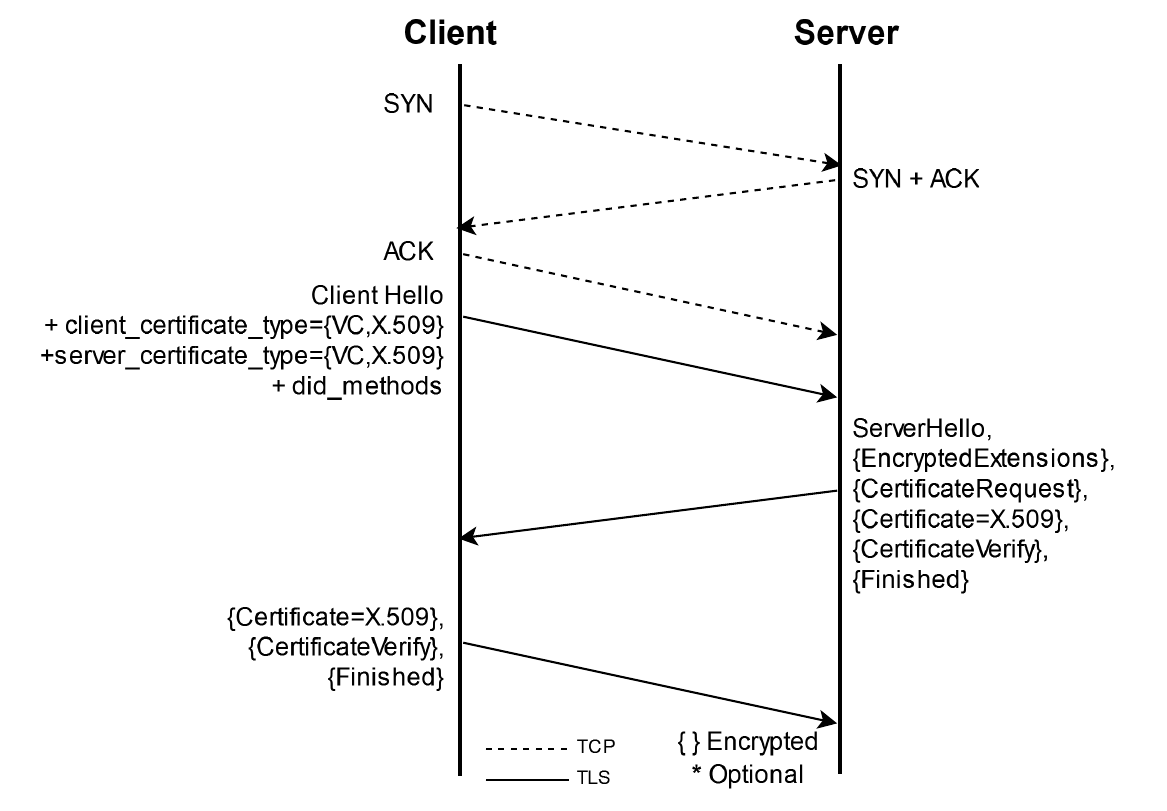}}
        \caption{}
        \label{fig:fallback-tls-hs}
    \end{subfigure}
\caption{Flow of messages in (a) a hybrid handshake where the client uses an X.509 certificate type and the server a \texttt{VC} certificate type and (b) a fallback to the original TLS 1.3 handshake protocol.}
\label{fig:hybrid-and-fallback-hs}
\end{figure*}

\subsection{Handshake with VC} 
\label{sec:vc-handshake}

Our main objectives are to (\emph{i}) extend the protocol to allow a server (and optionally a client) to authenticate with the peer using VCs, (\emph{ii}) preserve the interoperability with X.509 public key certificates, \emph{i.e.} support hybrid handshake with X.509 certificates and VCs and fallback to the original handshake, and (\emph{iii}) retain all the security features of TLS 1.3. 

We achieve these objectives by leveraging the \texttt{client\_certificate\_type} and \texttt{server\_certificate\_type}  extensions introduced by RFC-7250~\cite{rfc7250} in the TLS 1.3 handshake protocol. These two extensions are used to negotiate certificate types different than X.509 certificates. We added the \texttt{VC} certificate type to the already existing \texttt{X.509} and \texttt{RawPublicKey} types to enable the use of VCs. The exchange of the messages in the TLS handshake with the new certificate type is represented in Fig.\ref{fig:vc-tls-hs}.
The client sends the two extensions in the \texttt{ClientHello} indicating the support for the \texttt{VC} certificate type for both client and server authentication. The server sends the two extensions in the \texttt{EncryptedExtensions} message, selecting the \texttt{VC} type from the list presented by the client for both the client and the server.

We also added a new extension called \texttt{did\_methods}. The extension indicates the DID methods a TLS endpoint supports (\emph{i.e.}, the ledgers it can interact with) to resolve the peer's DID document. The structure of the extension retains the same syntax of all the other extensions as in RFC-8446~\cite{rfc8446} and it is shown below:
\vspace{-1mm}
\begin{verbatim}
enum {
   name0(0),
   name1(1),
   name2(2),
   ..
   (65535)
} DIDMethod
struct {
   DIDMethod did_methods<2..2^16-2>
} DIDMethodList
\end{verbatim}

The list of existing DID methods is currently maintained by the W3C in~\cite{DID-registry}. Each DID method is expressed in the form of a string. We propose a DIDMethod enum to map these strings into integer values for easiness of implementation. The \texttt{extension\_data} field carries the DIDMethodList structure.

The \texttt{did\_methods} is sent in the \texttt{ClientHello} and \texttt{EncryptedExtensions} messages. A client that sends the \texttt{server\_certificate\_type} extension containing the \texttt{VC} certificate type must also send the \texttt{did\_methods} extension. Similarly, a server that selects \texttt{VC} as the certificate type for the \texttt{client\_certificate\_type} extension must also send the \texttt{did\_methods} extension. A server must send a list of DID Methods both client and server share if it receives the \texttt{did\_methods} extension from the client. 

During the authentication phase, the server must check that the DID in its VC belongs to one of the DLTs present in the list of DID methods received by the peer, otherwise it must abort the session. 

Then, the server sends the VC in the \texttt{Certificate} message and proves its identity through the \texttt{CertificateVerify} message. The signature is computed with the $sk$ of the server over the hash of all the handshake messages exchanged up to that point. Its content includes the \texttt{ClientHello.random} and the \texttt{ServerHello.random}, which provide randomness, and the VC of the server. This replicates the design principle of the current SSI model that we discussed in section~\ref{sec:ssi} where a Holder wraps the VC and a nonce received from the Verifier into a VP, signs it and sends it back to the latter for authentication. 

The use of VCs reduces the total number of signatures in the handshake. VC Issuers are not organized hierarchically. The server must verify only the signature of the Issuer in the \texttt{Certificate} message. When using X.509 certificates, it must validate the certificate chain up to the Root CA. This is considered an advantage in the case of resource constrained devices, as it can improve performance while maintaining compliance with the VP concept. The server computes the \texttt{Finished} message as in the original TLS 1.3 handshake protocol.

Upon receiving \texttt{CertificateVerify} message, the client checks that the VC follows the scheme specified in the \texttt{@context} field, checks the validity of the VC metadata, resolves the DID of the Issuer and verifies its signature on the VC, and then extracts the server DID from the \verb+credentialSubject+ field of the VC and resolves the server DID to retrieve the server public key from the ledger. Finally, the client verifies the signature in the \texttt{CertificateVerify} message. Note that in the context of IoT systems, the number of available Issuer(s) is expected to be limited (one very often), therefore both client and server could securely maintain the public key of Issuer(s) they trust to avoid the DID resolution. Finally, the client verifies the \texttt{Finished} message. Then, it authenticates with the server in the same way.

Note that the use of \texttt{client\_certificate\_type} and \texttt{server\_certificate\_type} extensions in our design allows the two endpoints to authenticate with different certificate types. A client can authenticate with an X.509 certificate while the server can use its VC, as depicted in Fig.~\ref{fig:client-x509-server-vc-hs}, and vice versa. 
In addition, in the case a server does not support the \texttt{client\_certificate\_type} and the \texttt{server\_certificate\_type} extensions, it can always fall back to the original handshake, as depicted in Fig.~\ref{fig:fallback-tls-hs}. Through the use of these two extensions, the process towards full adoption of SSI can be incremental.  

\section{Security and Privacy Analysis}
\label{sec:security}

\subsection{Security}

The TLS handshake protocol with \texttt{VC} certificate type maintains the same security features of the original handshake, as the design remains the same. The only addition is the \texttt{did\_methods} extension. When the extension is sent in the \texttt{ClientHello} message, it benefits from the integrity and authentication properties. When it is sent in the \texttt{CertificateRequest} message, it benefits also from confidentiality.  

Most of the security discussion on our design focuses on the interaction with the DLT node for peer authentication. In the case of a DID resolution performed in clear, an active man-in-the-middle (MiTM) could impersonate the DLT node, forge a DID document containing the authenticating endpoint's DID, associate it with a key pair that he owns, and then return it to the DID resolver. Thus, the attacker is able to compute a valid \texttt{CertificateVerify} message by possessing the long term private key. In practice, the MiTM attacker breaks in transit the immutability feature of the DLT (\emph{i.e.}, the RoT for identity public keys). To mitigate this risk, an endpoint can establish a secure channel towards the DLT node, either through the Internet Protocol Security (IPSec) \cite{rfc6071, rfc7296, rfc4301} or TLS. They both provide authentication of the node and integrity and confidentiality of the data exchanged, but they come with an additional overhead. However, there are some differences among the two approaches. 

An IPSec channel works at network layer, and it is established before the TLS handshake takes place and can be reused until it expires. The moment the endpoint performs the DID Resolve, the messages are immediately ciphered and exchanged. An IPSec channel can only be established in case of full control of a DLT node, because additional software must be installed on it. In the case of IoT systems we consider the DLT nodes to be internal to the system and so under full control of the administrator. IPSec employs the Internet Key Exchange (IKE) \cite{rfc7296} protocol to authenticate the parties. In IKE authentication must be mutual. One option could be the usage of X.509 certificates, but it is not convenient because all the devices in the system should have an X.509 certificate and so they can directly run the original TLS 1.3 handshake protocol, making VCs worthless. Raw Public Keys (RPKs) \cite{rfc7250} represents a better option for authentication in this case. Every IoT node must save locally the public key of each DLT node and each DLT node must save locally the key of each IoT node. The number of DLT nodes within an IoT system is expected to be very low (\emph{i.e.}, one or a couple of nodes) with respect to the total number of IoT and edge nodes in the system, so the number of public keys stored by the latter is minimal. Only the DLT nodes should have the full list of public keys of the devices in the IoT system. One could argue that devices in the IoT system  could communicate securely by using directly RPKs for authentication in the TLS handshake, but this means that each IoT and edge node should store locally the public keys of all the other nodes in the system and this is again critical in terms of cost associated to identity management.    

A TLS channel towards the DLT node, instead, is established during the handshake between two IoT nodes right before the DID Resolve. Even if only server-side authentication is required in this case, this approach could slow down the handshake between IoT nodes. In order to reduce this overhead an endpoint could leverage the session resumption plus the zero round trip time (0-RTT) feature provided by the TLS 1.3 handshake protocol. 

In conclusion, although IPSec and TLS approaches provide the same security features for communication with the DLT node, we recommend the use of IPSec due to better performance in terms of handshake latency between IoT nodes, as the secure channel is established beforehand.

\subsection{Privacy}

Even though the \texttt{did\_methods} extension in the \texttt{ClientHello} is sent in clear no privacy issues arise as its content is not considered strictly confidential. 

In general terms, privacy issues can arise when the client resolves the server's DID on a public DLT node. The DLT node can monitor all the servers a client connects to. This problem disappears when DLT nodes are deployed as part of the IoT system itself.

\section{Implementation in OpenSSL}
\label{sec:implementation}

We have implemented the design presented in section \ref{sec:vc-handshake} in OpenSSL, a globally adopted open source cryptographic library, to experimentally evaluate it. OpenSSL is mainly written in C language and consists of two sub-libraries: \textit{ssl} that implements the TLS protocol and \textit{crypto} that supplies a wide variety of cryptographic operations. Internally, \textit{crypto} relies on loadable modules, called providers, that implement all the cryptographic logic.  

The main implementation goal is to minimize the changes to the OpenSSL sub-libraries to make smoother the contribution of the code upstream. In the end, our implementation consists of slight modifications to the \textit{ssl} module and no changes at all to the \textit{crypto} library.

In \textit{ssl} module we added the functions to construct and process the \texttt{did\_methods} extension, and the \texttt{VC} certificate type for the \texttt{client\_certificate\_type} and \texttt{server\_certificate\_type} extensions.

In \textit{crypto}, we treated the functions to deal with DIDs and VCs as cryptographic operations. We designed and implemented the \textit{ssi} provider to supply these functions to OpenSSL and use them in the TLS handshake protocol. To leave the \textit{crypto} library unchanged, we mapped these functionalities into four existing operations: 
\begin{itemize}
    \item \texttt{OSSL\_OP\_KEYMGMT}: manages the VC and DID document of an endpoint as an asymmetric key pair. The asymmetric key pair contains the VC in the public part and the DID document plus the identity private key as the private part. 
    \item \texttt{OSSL\_OP\_ENCODER}: to encode VCs and DID documents either in PEM or DER format;  
    \item \texttt{OSSL\_OP\_DECODER}: to decode VCs and DID documents from PEM or DER format;
    \item \texttt{OSSL\_OP\_SIGNATURE}: to construct and process the \texttt{CertificateVerify} message;
\end{itemize}

\begin{figure}[t]
    \centering
    \includegraphics[width=\columnwidth]{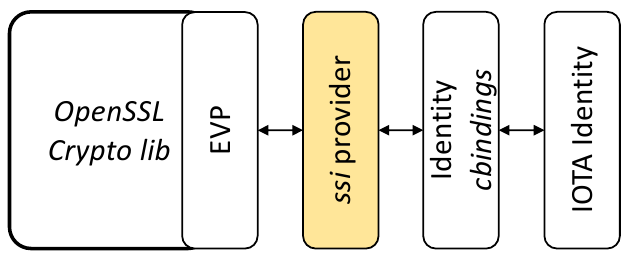}
    \caption{The high-level architecture of the overall solution.}
    \label{fig:ssi-provider}
\end{figure}

Fig.~\ref{fig:ssi-provider} shows the architecture of our overall solution.
The implementation of the \emph{ssi} provider operations relies on the identity-cbindings library. We developed the \emph{identity-cbindings} library as a wrapper of the IOTA Identity library \cite{identity}. The IOTA Identity library provides DID and VC-related functionalities in accordance to W3C specifications and the \emph{iota} method to work with the IOTA Tangle ~\cite{Tangle}. The purpose of the wrapper is to provide the \emph{ssi} provider with C language interfaces to communicate with the IOTA Identity library, which is written in Rust language. 

Before running the TLS handshake, both endpoints generate their self-sovereign identity by executing the OpenSSL \textit{genpkey} application with the \texttt{VC} option set. This application allows an endpoint to generate through our \textit{ssi} provider the identity key pair, a DID document and a VC calling the \texttt{OSSL\_OP\_KEYMGMT} operation. Then, it encodes these identity components in PEM format through the \texttt{OSSL\_OP\_ENCODING} operation.

The client and the server run the \textit{s\_client} and \textit{s\_server} applications respectively to start the TLS handshake. Both applications use the same cryptographic operations of the \textit{ssi} provider. At the beginning, these two applications load their identity and invoke the \texttt{OSSL\_OP\_DECODING} operation to decode them from PEM format and store them in IOTA Identity structures. 

During the TLS handshake, the endpoint invokes the  \texttt{OSSL\_OP\_ENCODING} to encode the VC in DER format and place it in the \texttt{Certificate} message, and the \texttt{OSSL\_OP\_SIGNATURE} to sign the \texttt{CertificateVerify} message with IOTA identity cryptographic functions. 
The peer processes the \texttt{Certificate} and \texttt{CertificateVerify} messages by calling, respectively, the \texttt{OSSL\_OP\_DECODER} operation to decode the VC received from the peer and store it in IOTA Identity structures, and the \texttt{OSSL\_OP\_SIGNATURE} operation to verify the peer's identity as in section \ref{sec:vc-handshake}. 

The open source implementation is available in \cite{openssl, openssl-ssi-provider, identity-cbindings}. 

\section{Performance Analysis}

\subsection{Experimental Setup}
\label{sec:setup}

\begin{figure}[t]
    \centering
    \includegraphics[width=\columnwidth]{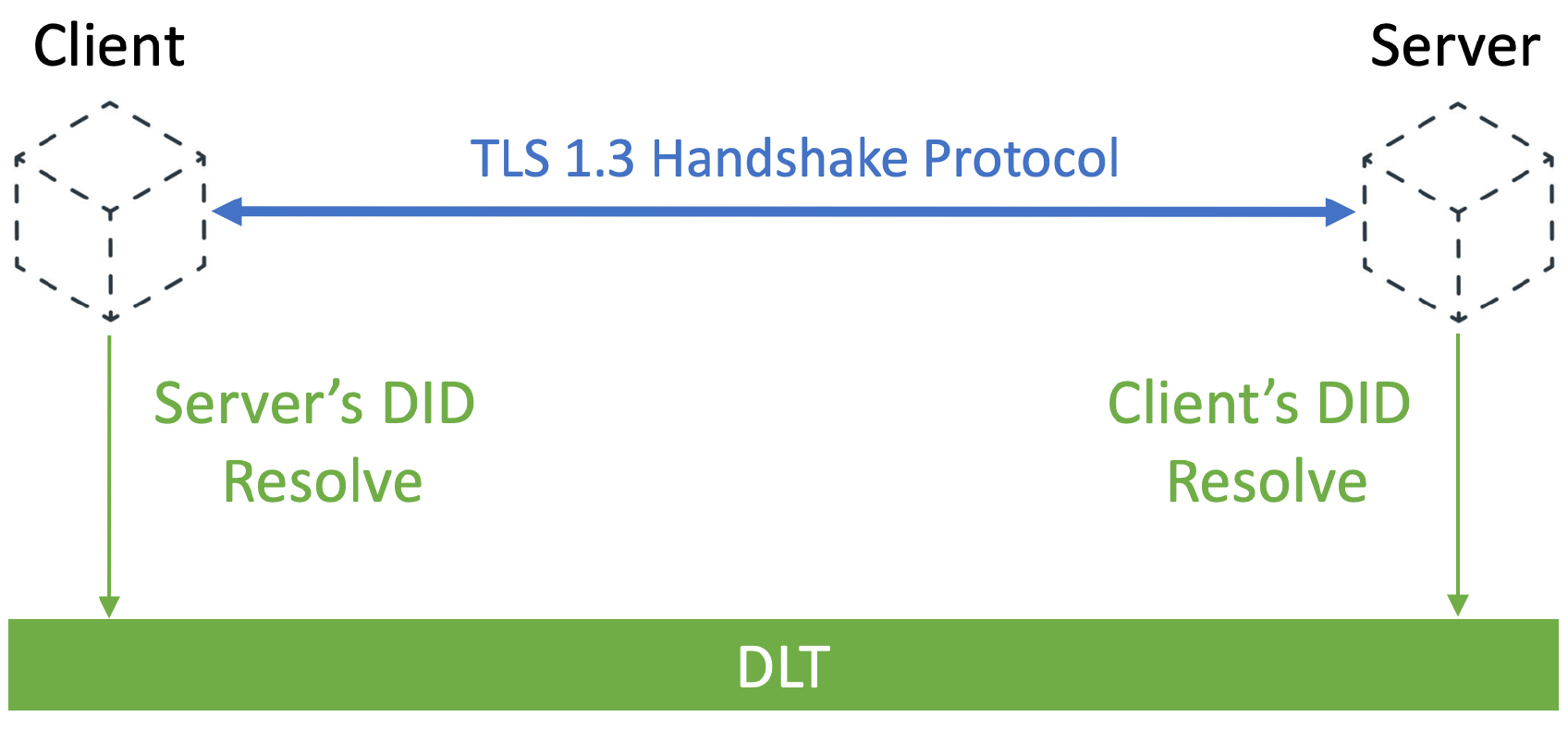}
    \caption{The experimental setup with two Raspberry Pi 4 Model B leveraging the IOTA Tangle DLT as the RoT for public keys.}
    \label{fig:setup}
\end{figure}

To assess the performance of the new handshake we have installed the modified version of OpenSSL on two Raspberry Pi's 4 (RPI) Model B equipped with a quad-core Cortex-A72 (ARM v8) SoC clocked at 1.8GHz, 4 GB of SDRAM, a Gigabit Ethernet interface, and 32-bit OS. The RPIs are connected in a client-server configuration and both have access to a DLT node to resolve the peer DID as depicted in Fig.~\ref{fig:setup}. Without loosing generality, the RPIs leverage the immutability feature of the public IOTA Tangle \cite{Tangle} and interact with it through an IOTA node operated in our lab. The TLS endpoints resolve the peer DID with an HyperText Transfer Protocol (HTTP) request served by the IOTA node.

We tested the TLS handshakes between the two target RPIs under three different  configurations for the connection between the RPIs and the IOTA node: (\textit{i}) HTTP over TCP, (\textit{ii}) HTTP over TCP with IPSec in endpoint-to-endpoint transport mode at layer 3, and (\textit{iii}) HTTPs with a TLS channel established on the fly during the handshake between the RPIs. In the IPSec scenario, the RPIs authenticate the IOTA node with an Ed25519 Raw Public Key, while in the TLS scenario the node employs a three-link X.509 certificate chain signed with Ed25519 keys.

In all tests, client and server adopt x25519 elliptic curve for Elliptic Curve Diffie-Hellman Ephemeral (ECDHE) key exchange and \texttt{TLS\_AES\_256\_GCM\_SHA384} cipher suite \cite{rfc8446}. 
We tested the performance of unilaterally authenticated handshake and mutually authenticated handshakes and compared them with the ones of the original handshake using a three-link X.509 certificate chain. We have only considered the Ed25519 signature algorithm, as it is the algorithm supported by the IOTA Identity library \cite{identity}. It relies on the cryptographic implementation of Ed25519-zebra library \cite{ed25519-zebra}.
We ran 1000 TLS handshakes for each configuration using \texttt{s\_client} and \texttt{s\_server} applications provided by OpenSSL to collect statistically relevant results about handshake size and latency. Client and server select randomly an X.509 certificate chain or a VC from a predefined large set at each run.

\subsection{Experimental results}
\label{results}

\subsubsection{handshake size} Table \ref{tab:hs-size} shows the number of bytes sent by the server to the client in a unilaterally authenticated handshake while using X.509 and \texttt{VC} certificates. The content, and so the number of bytes, in a VC or X.509 certificate can vary based on how much information they contain. Therefore, the total bytes column should be considered as indicative. Instead, the values about public key objects are fixed. In a handshake with \texttt{VC} certificate type, a server sends half of the bytes of public key objects than an original handshake does. 
Note that in a unilaterally authenticated handshake with the \texttt{VC} certificate type, only the client resolves the server DID. Conversely, in a mutually authenticated handshake also the server interacts with the DLT node. In the latter scenario, when the server establishes a TLS channel with the DLT node to resolve the client DID, it sends an additional 377 bytes. 
In general, endpoints may find more suitable the IPSec alternative as the secure channel gets established before the handshake and so fewer bytes will be exchanged over the network. In the case of mutually authenticated handshakes, the values about public key objects in Table \ref{tab:hs-size} double.

\begin{table}[t]
    \centering
    \begin{tabular}{lc|cccc}
        \toprule
        \multirow{2}{*}{} & \textbf{Total Bytes} & \multicolumn{3}{c}{\textbf{Public Key Objects}} \\
        {} & {} & pk & signature & tot \\
        \midrule
        \textbf{X.509} & 1023 & 2*32 & 3*64 & 256 & \\
        \textbf{VC} & 1331 & / & 2*64 & 128 & \\
        \bottomrule
    \end{tabular}
    \caption{Data sent by the server during an unilaterally authenticated TLS 1.3 handshake [bytes].}
    \label{tab:hs-size}
\end{table}

\begin{table}[t]
    \centering
    \begin{tabular}{ccc|cccc}
        \toprule
        {} & {} & \textbf{X.509} & \textbf{HTTP} & \textbf{HTTP over TCP} & \textbf{HTTPs} \\
        {} & {} & \textbf{} & \textbf{over TCP} & \textbf{with IPSec} & \textbf{} \\
        \midrule
        \multirow{2}{*}{\rotatebox[origin=c]{90}{\textbf{UNI}}} & Handshake & 13,9 & 25,5 & 26,4 & 31,2 \\ 
        {} & DID res. & & 8,2 & 9,2 & 15,3 \\
        \midrule
        \multirow{2}{*}{\rotatebox[origin=c]{90}{\textbf{MUT}}} & Handshake & 21 & 46,4 & 50 & 64,6 \\
        {} & DID res. & & 17,8 & 21,4 & 38 \\
        \bottomrule
    \end{tabular}
    \caption{Average latency of the overall TLS handshake and DID resolutions in unilaterally and mutually authenticated handshakes [ms].}
    \label{tab:unilateral-and-mutual-handshake}
\end{table}

\begin{table}[t]
    \centering
    \begin{tabular}{l|cc}
        \toprule
        \textbf{Server} & X.509 & VC\\
        \textbf{Client} & VC & X.509 \\
        \midrule
        Handshake & 41,6 & 40,7 \\
        \bottomrule
    \end{tabular}
    \caption{ Average latency of the overall TLS hybrid handshakes with IPSec connection to the DLT node [ms].}
    \label{tab:hybrid-handshake}
\end{table}

\subsubsection{unilaterally authenticated handshakes} the upper part of Table~\ref{tab:unilateral-and-mutual-handshake} shows the average latency of a unilaterally authenticated handshake using either an X.509 or a \texttt{VC} certificate type. 

The original handshake is overall faster than all the handshakes with \texttt{VC} certificate type, even though the results are of the same order of magnitude.
The overhead is not a matter of cryptographic operations, both OpenSSL and IOTA Identity spend the same time to sign (on average 0,2 ms) and verify a message (on average 0,6 ms). 
The latency difference between the handshake with the \texttt{VC} certificate type and the original is mainly due to the two DID Resolve operations during the VC verification (\emph{i.e.}, issuer's DID and peer's DID). The two resolutions account for 71\% with HTTP over TCP, 73\% with HTTP over TCP with IPsec, and 88\% with HTTPs of the total difference. 
The configuration with HTTP over TCP to the DLT node provides the fastest handshake, but does not provide any security. Adding IPSec in transport mode to the previous configuration provides fast communication and end-to-end security with a 
small overhead, as the IPSec setup takes place beforehand. The HTTPs solution slows down the handshake between the target nodes due to the need to open another TLS connection with the DLT node to resolve the DIDs.

Overall, the experimental results show that the IPSec approach offers the best trade-off between security and performance. However, the performances in Table~\ref{tab:unilateral-and-mutual-handshake} represents the worst case. In fact, in an IoT system we suppose there is only one Issuer and its DID and public key can be stored on each IoT node halving the number of required DID Resolves, hence the overall latency of a handshake between the target IoT nodes. The DID Resolve operation would drop from 9,2 ms to 5,2 ms and the TLS handshake latency from 26,4 ms to 22,4 ms on average.

\subsubsection{mutually authenticated handshakes}
the lower part of Table~\ref{tab:unilateral-and-mutual-handshake} shows the average latency of a mutually authenticated handshake using either X.509 or \texttt{VC} certificate types. The handshakes are slightly slower as both peers authenticate. 
The same considerations made for the unilaterally authenticated handshakes hold in this case. The use of IPSec to protect the communications with the DLT node remains the best approach, as it does not present substantial degradation of performance compared to the use of HTTP over TCP. Again, these results represent the worst case. 
In fact, the DID resolution latency includes four DID Resolve, two performed by the client and two by the server. By storing the Issuer's DID and public keys locally in the target IoT nodes the overall DID Resolve latency drops from 21,4 ms to 12,2 ms on average and the average handshake latency decreases from 50 ms to 40,8 ms on average, improving performance of about 18\%.

\subsubsection{hybrid handshakes and fallback} 
Table~\ref{tab:hybrid-handshake} shows the average latency of hybrid TLS handshakes between the target IoT nodes. We tested the TLS handshakes while protecting the communication with the DLT node by means of IPSec.
As expected, the average latency are almost equal in both cases, since the sequence of operations performed are the same. 
In addition, we tested the fallback to the original handshake to cover the case a server does not implement RFC-7250~\cite{rfc7250} or does not recognize the new \texttt{VC} certificate type. We installed two different versions of OpenSSL on the two IoT nodes; one with our implementation on the client and an original version on the server. The functional tests were successful. The average handshake latency remains the same as in the original handshake case.

\section{Conclusion and Future Work}

To support and facilitate the adoption of SSI in large-scale IoT systems, we have taken a step forward in the design and implementation of a TLS 1.3 handshake protocol that enables the use of VCs while maintaining full compliance with RFC-8446 and preserving all the security features of TLS 1.3.
The improvement over our previous work lies in the handshake design, which now only uses messages already defined for TLS 1.3 and a new extension. The design has an incredibly positive impact on the implementation, as we made minimal changes to the OpenSSL library and relied mostly on our external \textit{ssi} provider to handle VC and DID-related operations. 
Furthermore, the experimental results prove the feasibility of our design and show comparable performance to the original solution based on PKI and X.509 certificates. These results pave the way for the adoption of SSI in large-scale IoT systems, with a clear benefit in terms of reducing the cost of identity management. Future work will focus on further optimizing the implementation to further improve performance.    

\section*{Acknowledgment}
This work has been developed within the SEDIMARK project (\url{https://sedimark.eu}), funded by the European Union under the Horizon Europe framework programme [grant agreement no. 101070074].

\bibliographystyle{IEEEtran}
\bibliography{biblio}

\end{document}